\documentclass[aps,prb,twocolumn,groupedaddress]{revtex4}

\usepackage{graphicx}
\usepackage{xspace}

\providecommand{\ie}{i.e.,\xspace}
\providecommand{\kb}{$\overline{{K}}$\xspace}
\providecommand{\kg}{$\overline{{K}}-\overline{\Gamma}$\xspace}
\providecommand{\km}{$\overline{{K}}-\overline{{M}}$\xspace}
\providecommand{\epc}{electron-phonon coupling\xspace}
\providecommand{\se}{self-energy\xspace}

\begin{document}


\title{Electron-phonon coupling in potassium-doped graphene: Angle-resolved photoemission spectroscopy}


\author{M. Bianchi}
\affiliation{Physics Department, University of Trieste, Via Valerio 2,
I-34127 Trieste, Italy}
\author{E.~D.~L.~Rienks}
\affiliation{Institute for Storage Ring Facilities and Interdisciplinary Nanoscience Center (iNANO), University of Aarhus, 8000 Aarhus C, Denmark}
\author{S. Lizzit}
\affiliation{Sincrotrone Trieste, S.S. 14 Km. 163.5, I-34149 Trieste, Italy}
\author{A. Baraldi}
\affiliation{Physics Department and CENMAT, University of Trieste, Via Valerio 2,
I-34127 Trieste, Italy}
\affiliation{Laboratorio TASC INFM-CNR, S.S. 14 Km 163.5, I-34012 Trieste, Italy}
\author{R. Balog}
\affiliation{Institute for Storage Ring Facilities and Interdisciplinary Nanoscience Center (iNANO), University of Aarhus, 8000 Aarhus C, Denmark}
\author{L. Hornek\ae r}
\affiliation{Institute for Storage Ring Facilities and Interdisciplinary Nanoscience Center (iNANO), University of Aarhus, 8000 Aarhus C, Denmark}
\author{Ph. Hofmann}
\affiliation{Institute for Storage Ring Facilities and Interdisciplinary Nanoscience Center (iNANO), University of Aarhus, 8000 Aarhus C, Denmark}
\email[]{philip@phys.au.dk}
\homepage[]{http://www.phys.au.dk/~philip/}



\date{\today}

\begin{abstract}
The electron-phonon coupling in potassium-doped graphene on Ir(111)
is studied via the renormalization of the $\pi ^{\ast}$ band near
the Fermi level, using angle-resolved photoemission spectroscopy.
The renormalization  is found to be fairly weak and almost isotropic,
with a mass enhancement parameter of $\lambda=$ 0.28(6) for both the \km
and  the \kg direction. These results are found to agree
well with recent first principles calculations.
\end{abstract}

\pacs{73.22.Pr,79.60.-i,71.38.Cn,73.20.At}

\maketitle


Many-body effects in graphene have recently attracted
considerable attention, mainly because of graphene's perceived role as a
future electronics material, but also because graphene is highly suited as
a fundamental test case for the investigation of many body effects 
\cite{Bostwick:2009,Bostwick:2007,Bostwick:2007b,McChesney:2007,McChesney:2008,Basko:2008,Lazzeri:2008,Park:2008b,Park:2009}. In
particular, the structural and electronic simplicity of graphene
seems to imply that it is accessible to both first-principles
calculations and angle-resolved photoemission (ARPES).
Other advantages are that high-quality graphene can be
prepared and doped by an electric field or by adsorption.

This paper is concerned with the electron-phonon interaction
in electron-doped graphene. Despite the above-mentioned apparent
advantages of graphene to study many-body interactions, there has
been considerable dispute on the strength of the electron-phonon
coupling in both experimental and theoretical studies. From the
experimental side, there are several possible causes for this. First
of all, experiments have not been performed on free-standing graphene
and it is not easy to evaluate the role of the substrate. More
importantly, determining the electron-phonon self-energy from the
dispersion's renormalization is not simple, after all, since the unrenormalized (bare)
dispersion is also influenced by many-body effects at higher
energies \cite{Bostwick:2007,Bostwick:2007b}. The usual approach
of using a linear bare dispersion fails for graphene \cite{Park:2008b},
a fact which has lead to unrealistic results in early studies
\cite{McChesney:2007}. Temperature-dependent studies, usually a
stable alternative to determine the coupling strength in the form of the
electron-phonon mass enhancement parameter $\lambda$, are of limited
value because of graphene's exceptionally high Debye temperature.
At this point, rather different experimental results for the
size of $\lambda$ and its
variation over the Fermi surface have been reported for 
graphene and graphite intercalation compounds
\cite{McChesney:2007,McChesney:2008,Valla:2009,Gruneis:2009}.

Theoretical results generally point towards a rather weak coupling
but the absolute values of $\lambda$ vary greatly between different
calculations \cite{Calandra:2007,Park:2008b}. Moreover, it has been
pointed out that the local density approximation might not be able
to predict the \epc correctly \cite{Basko:2008,Lazzeri:2008}. On
the positive side, a very recent calculation of the expected linewidth
in ARPES, including both electron-electron and electron-phonon
scattering effects gives good agreement with experimental data
for doped graphene on SiC, at
least in the \km direction for states near the Fermi energy ($E_F$)
\cite{Park:2009}.

In this paper, we analyze the electron-phonon coupling strength for
electron-doped graphene grown on Ir(111). This approach has the
potential advantage of starting with a graphene layer which is
relatively well decoupled from the substrate and only very weakly
doped when clean. We use a stable method to determine $\lambda$
despite the unknown bare dispersion. We find a moderate $\lambda$
value which varies little over the Fermi surface, a scenario that
agrees well with a recent theoretical prediction \cite{Park:2008b}.
We compare our results to other works using graphene on SiC as a
starting material  \cite{McChesney:2007,Bostwick:2007,Bostwick:2007b}
as well as the KC$_8$ and CaC$_6$ intercalation compounds
\cite{Valla:2009,Gruneis:2009}. In KC$_8$, and probably also in
CaC$_6$, the graphene sheets are effectively decoupled because of
the increased distance between them after alkali
intercalation\cite{Gruneis:2009b}.

%

Experiments were performed at the SGM3 beamline of the ASTRID
synchrotron radiation facility at  \AA rhus  University
\cite{Hoffmann:2004}. The endstation is equipped with a 150~mm
hemispherical electron energy analyzer (Specs). The data were taken
with a photon energy of 45~eV, and at a sample temperature of 70~K.
The total energy and  $k$ resolution amounted to 28~meV and
0.01~\AA$^{-1}$, respectively.

The graphene film was prepared on Ir(111) using a standard
recipe~\cite{Pletikosic:2009,Lacovig:2009}.  The cleanness of the
Ir substrate before graphene formation was monitored using the 4$f$
surface core level shift, and the quality of the graphene film was
checked by means of Low Energy Electron Diffraction (LEED) and
ARPES, which showed the characteristic mini-gaps close to the \kb
point \cite{Pletikosic:2009}.  Potassium was evaporated from a
commercial getter source (SAES).  The data reported here were taken
in a situation in which LEED showed a $(2 \times 2)$ superstructure.
This corresponds to the KC$_8$ phase in an alkali metal / graphite
intercalation compound.  At this doping level, the Dirac point is
shifted to a binding energy of 1.29~eV and the electron concentration
is $\approx 1\times 10^{14}$~cm$^{-2}$ or, equivalently, the doping
level is 0.054 extra electrons per graphene unit cell.

\begin{figure}
\includegraphics[width=3in]{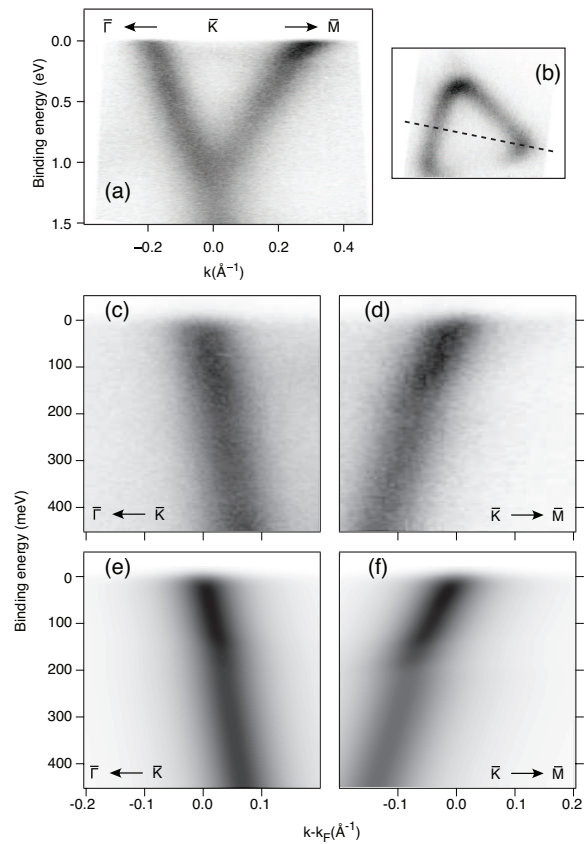}%
\caption{Photoemission intensity around the \kb point shown as
different cuts through a three-dimensional data set. (a) large-scale
dispersion through \kb. (b) Fermi surface around \kb with the
dashed line indicating the cut shown in (a). More detailed Fermi energy 
crossings along the \kg (c) and \km (d) directions. (e) and (f)
show the simulated photoemission intensity corresponding to the
data in (c) and (d), respectively. 
\label{fig:disp}}
\end{figure}

An area of two-dimensional reciprocal space around \kb was sampled
and a resulting dispersion and the Fermi contour are shown in Fig.
\ref{fig:disp} (a) and (b), respectively. Detailed dispersions along
the \km and \kg directions were extracted and shown in Fig.
\ref{fig:disp}(c) and (d), respectively.  The dispersion along the
\kg direction in (c) is a sum of four parallel, closely spaced slices. Along
\km only a single cut has been used.  Averaging over several parallel
cuts in this direction gives rise to broadening, as expected because
of the sharp corner of the Fermi surface in this direction. The
dispersions along \km and \kg both show a pronounced renormalization
of the bands, as evident from the kink in the dispersions around a
binding energy of $\approx180$~meV. The apparent strength of the
kink is greater in the \km direction than in the \kg direction.
Qualitatively, these results are in good agreement with previous
findings in similar systems
\cite{McChesney:2007,McChesney:2008,Valla:2009,Gruneis:2009}.

Due to the three-fold symmetry of the Fermi surface, several
equivalent cuts can be made. For the following analysis we have
focused on the directions with the highest  signal to noise ratio
and smallest contribution of the underlying Ir surface states. The
results from other directions are consistent with the reported
findings and have been used to estimate the uncertainty of the
obtained coupling strengths.

The detailed renormalized dispersions were extracted by fitting
Lorentzian lines with a linear background to the momentum distribution
curves (MDCs), i.e. to constant energy profiles. The fits were of
good quality even though, strictly spoken, MDC cuts through the
spectral function can only be described by Lorentzian lines in the case
of a linear dispersion. This procedure also gives the energy-dependent
full width at half maximum (FWHM) of the MDC peaks.

In the next step of the analysis, the \se $\Sigma$ has to be
extracted from the data. $\Sigma$ is a complex quantity and its
real and imaginary parts are related via a Kramers-Kronig transformation.
The measured spectral function contains information about both the
real part $\Sigma'$ and the imaginary part $\Sigma''$ of the
self-energy via the size of the renormalization and the linewidth
of the MDCs, respectively \cite{Hofmann:2009b}. Extracting  $\Sigma''$
from the MDC width suffers from the problem that the width is usually
rather noisy and that, strictly, $\Sigma''$ can only be determined
in the case of a linear bare dispersion. Using the renormalized
dispersion to extract $\Sigma'$, on the other hand, requires the
knowledge of the bare dispersion. In the present case, both
restrictions pose formidable problems but several approaches have
been suggested in order to overcome them, such as measuring the
dispersion at high temperatures  \cite{Kirkegaard:2005} or using a
self-consistent procedure to extract $\Sigma$ \cite{Kordyuk:2005}.
We employ a procedure which is similar to the latter approach.

In the first step of this analysis, a model for the measured spectral
function is calculated from initial guesses for the bare dispersion
and $\Sigma$. The bare dispersion is assumed to be a second order polynomial.
$\Sigma$ is calculated from an Eliashberg coupling function
$\alpha^2F(\omega)$, consisting of the contributions of 5 Einstein
oscillators with energies that are evenly distributed over the
energy range from 21~meV to $\omega_{\mathrm{max}}=190$~meV.  The
oscillators, with energy $\omega_i$ and a coupling strength value
$\lambda_i$, have a width of 1~meV.  From $\alpha^2F(\omega)$, it
is straightforward to calculate $\Sigma''$ (and thus also $\Sigma'$)
through \begin{equation} \Sigma''(\omega)=
\pi\int_{0}^{\omega_{\mathrm{max}}} \alpha^2 F(\omega') [1 -
f(\omega-\omega') + f(\omega+\omega') + 2n(\omega') ] d \omega',
\label{life-qusi} \end{equation} where $n$ and $f$ are Bose and
Fermi functions, respectively. Note that the simple, discrete
character of the Eliashberg function has only a  minor effect on
$\Sigma''$.  Indeed, reasonable fits of $\Sigma''$ can sometimes
be achieved with even simpler models \cite{Valla:2009}. Adding further
Einstein oscillators in the present case would render them redundant in view of
the experimental energy resolution.

We add an offset to $\Sigma''$ in order to account for energy-independent
defect scattering. As we are only concerned with a small energy
window close to $E_F$, we ignore the effect of electron-electron
scattering altogether.  The model thus contains 9 variable parameters:
3 governing the bare-particle dispersion, 5 values of  $\lambda_i$
and one offset for $\Sigma''$.  Using these, the spectral function can
be calculated from the bare dispersion and $\Sigma$. The result is
multiplied with a Fermi function and convoluted with the experimental
energy and $k$ resolution functions.

In a second step, the MDC position and width are extracted from
this \emph{simulated} spectral function in the same way as from the
experiment.   The MDC peak positions and widths of experiment and
simulation are compared using a combined $\chi^2$  for position and
width. The model parameters are then optimized using a steepest-descent
algorithm.

\begin{figure}
\includegraphics[width=3in]{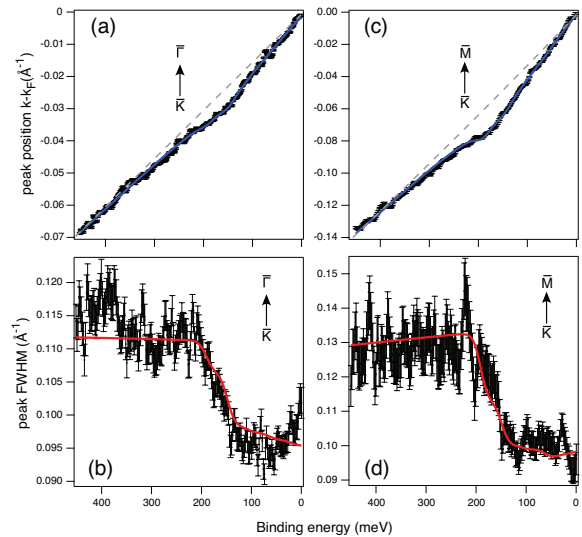}%
\caption{(Color online) Detailed comparison between the measured
spectral functions and those found in the iterative optimization
procedure outlined here. (a)  and (c)  peak position in MDC cuts
together with the bare dispersion obtained from the fit (dashed line);
(b) and (d) peak full width at half maximum. Points with error bars
represent data, solid lines the fit.  \label{fig:fitres}}
\end{figure}

\begin{figure}
\includegraphics[width=3.5in]{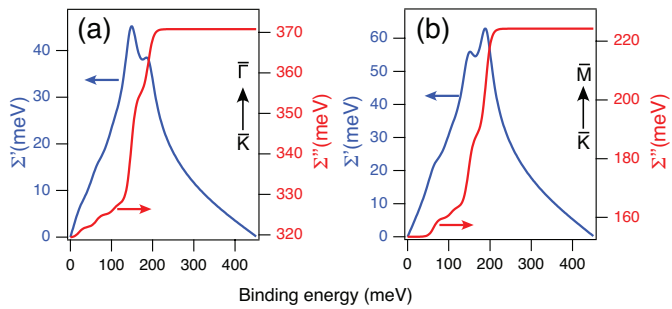}%
\caption{(Color online) Optimized self-energies (real part $\Sigma'$ and imaginary part $\Sigma''$) for
the \kg (a) and \km (b) directions. 
\label{fig:eself}}
\end{figure}

The results of this fitting procedure are given in Fig. \ref{fig:fitres} and \ref{fig:eself}. 
Fig. \ref{fig:fitres} shows the agreement between data and simulation as well
as the bare dispersion obtained from the fit. Fig.
\ref{fig:eself}  gives the resulting $\Sigma$.  The level of agreement in Fig. \ref{fig:fitres} is
clearly satisfactory given the simplicity of the model.  The \se
obtained from the procedure reveals the expected contributions
caused primarily by the high energy optical phonon modes 
and, less strongly, lower-lying modes, consistent with the literature
\cite{McChesney:2007,McChesney:2008,Valla:2009,Gruneis:2009}.  The
mass enhancement parameter $\lambda$ can now be calculated as the
first reciprocal moment of $\alpha^2F(\omega)$. We find the same
$\lambda$ value of 0.28(6)  for the \km and \kg
directions. 

It is quite remarkable that the obtained $\lambda$ values are so
similar for the two directions despite of the apparently stronger
renormalization along \km. The reason is the downward curvature of
the band in this direction as it approaches the van Hove singularity,
as noted earlier \cite{Park:2008b}. This curvature also leads to
counter-intuitive slight increase of the linewidth upon approaching
the Fermi level, best observed between 450 and 200~meV in Fig.
\ref{fig:fitres}(d) and also reported for similar compounds
\cite{Valla:2009}. An energy dependence of the linewidth in
regions with a rapidly changing density of states is, of course,
expected and has been found for other systems \cite{Gayone:2003}.
The similarity of the $\lambda$ values can already be guessed from
inspecting $\Sigma''$ alone. The absolute value of the two curves
in Fig. \ref{fig:eself} is different but the increase in  $\Sigma''$
due to electron-phonon coupling is very similar. Note that there
is no \emph{a priori} reason to assume anisotropic $\lambda$ values.
If the coupling to very low-energy acoustic phonons (\ie with
$\mathbf{q}$ vectors significantly shorter than the size of the \kb
Fermi contour) were relevant, one could expect shorter lifetimes
around the \km direction, as more states with similar energy
are close by. But as coupling to high energy optical phonons is
obviously dominant and intervalley scattering is thought to be
important for electron-doped graphene \cite{Valla:2009,Gruneis:2009},
this simple argument does not hold.

We note that the outlined approach for determining $\Sigma$ and
$\lambda$ has several advantages over other methods. First of all,
it does not rely on any knowledge of the bare dispersion, as long 
as it is locally parabolic. In the
case of graphene, this is crucial, because the strong non-linearity
in the bare dispersion close to the van Hove singularity does not
only upset a simple determination of $\Sigma'$, it also makes it
impossible to relate the MDC linewidth to $\Sigma''$ through the
usual relation that $\Sigma''= v \,\mathrm{FWHM} /2$, where $v$ is
the group velocity of the state and FWHM the MDC full width at
half-maximum. As mentioned before, this effect is most pronounced
along \km.

The data analysis approach also has the advantage of built-in
consistency between $\Sigma'$ and $\Sigma''$ at the expense
of  assuming particle-hole symmetry for the transformation from
$\Sigma''$ to $\Sigma'$. Also, it permits sufficient variation of
$\Sigma$, while describing the system with a manageable number of
parameters. Indeed, the idea of using an Eliashberg function
constructed with a small number of Einstein oscillators is simple
and stable and one might wonder why it works so well. The reason
is that one does not fit the actual Eliashberg function but $\Sigma$,
or rather the observed dispersion, which contains only a
temperature-broadened integral over the $\alpha^2F(\omega)$. Details
in the latter become insignificant but the determination of the
coupling strength is stable. Finally, the procedure includes the
band distortion caused by the finite resolution of the spectrometer.
In particularly the energy resolution is well-known to distort the
dispersion close to $E_F$, leading to an error in the band slope \cite{Kirkegaard:2005}. For the
determination of $\lambda$ this is obviously very significant because
for $T=0$, $\lambda$ is equal to the slope of $\Sigma'$ at $E_F$.

We compare our results to $\lambda$ values obtained for graphene
and graphite with similar electron doping.
For KC$_8$  it has been
recently shown that the very intercalation of K atoms also leads
to an effective reduction of the system's dimensions from three to
two, caused by the larger distance between the graphene sheets,
such that a direct comparison to graphene is possible \cite{Gruneis:2009}.

The analysis of the electron-phonon coupling for electron-doped
graphene grown epitaxially on SiC has given rather similar $\lambda$
values as reported here
\cite{McChesney:2007,Bostwick:2007,Bostwick:2007b}. An early study
had reported a rather large anisotropy, especially at even higher
doping levels, which was most probably caused by the unrealistic
assumption of a linear bare dispersion \cite{Park:2008b}. On KC$_8$,
on the other hand, $\lambda$ was found to be rather anisotropic and
also stronger than here \cite{Gruneis:2009}. The reason for this
is unclear and could lie in the different approach to data analysis
or in the fact that having K atoms \emph{in between} graphene sheets
leads to markedly different phonon dispersion, even though the
electronic properties of the individual graphene sheets are similar
to free graphene. Our results agree well with a recent
first principles calculations of the electron-phonon coupling in
electron-doped graphene \cite{Park:2008b,Park:2009} which predicts
an almost isotropic \epc of similar strength.

Finally we compare the absolute MDC linewidth close to the Fermi
level to that in similar systems in order to assess the role of
defect scattering in this system. The linewidth found here is roughly
0.095~\AA$^{-1}$ which is quite similar to data from
KC$_8$\cite{Gruneis:2009} but significantly broader than for
alkali-doped graphene on SiC \cite{Bostwick:2007}. It has to be
noted, however, that the graphene/SiC data in Ref.  \cite{Bostwick:2007} have
been taken at lower doping levels and MDCs taken for the highest
coverages suggest that there is an increase of linewidth in this
regime. A plausible explanation for an increased linewidth for
graphene on Ir(111) is the fact that the resulting superstructure
is incommensurate \cite{Diaye:2008}, implying a loss of translational
symmetry for the combined system and a presumably increased scattering
rate, even if the coupling between substrate and graphene is small.
We also note that the adsorption of alkali metals on graphene/SiC
and graphene/Ir(111) appears to be different. On SiC it is possible
to continuously vary the doping level whereas this cannot be done
for Ir(111). Here the potassium adsorbates assemble in islands with
$(2 \times 2)$ periodicity and at lower coverages two $\pi$ bands
are observed, one similar to the doped case reported here and one
similar to clean graphene on Ir(111) \cite{Bianchi_tbp}. Finally,
it is questionable if a comparison of the coupling strengths and
linewidths for different doping levels is at all meaningful. Valla
\emph{et al.} have evoked a dynamical intervalley nesting effect
to explain the observed coupling on CaC$_6$ and this effect implies
a strong doping dependence of the coupling \cite{Valla:2009}.

In conclusion, we have determined the electron-phonon coupling
strength for a layer of alkali-doped graphene on Ir(111) in the \kg
and \km directions. We find an almost isotropic scattering strength
consistent with recent first-principles calculations. The approach
used here takes the non-linearity of the bare dispersion and the
finite experimental resolution into account and uses a simple model
for the Eliashberg function which allows a stable, reliable and
self-consistent analysis of the measured spectral function.

We thank Marko Kralj for stimulating disussions.
This work has been supported by Lundbeck foundation, the Danish
National Research Council and the University of Trieste.  MB
acknowledges financial support from the University of Trieste within
the program of Student International Mobility 2007/2008.

\end{document}